\documentclass[a4paper]{jpconf}
\usepackage{graphicx}

\usepackage{ulem} 
\usepackage{lipsum}
\usepackage{physics}
\usepackage{listings}
\usepackage{color}
\definecolor{mygreen}{rgb}{0,0.5,0}
\usepackage{float}
\usepackage{hyperref}
\usepackage{array}

\newcommand{\myvector}[1]{\ensuremath{\begin{pmatrix}#1\end{pmatrix}}}
\definecolor{taha1}{rgb}{1,0,0}

\definecolor{taha2}{rgb}{0,0.6,0}

\definecolor{taha3}{rgb}{0,0,1}

\definecolor{taha4}{rgb}{1,0,1}

\usepackage{subcaption}

\usepackage{cite}

\begin{document}
\title{Compiling single-qubit braiding gate for Fibonacci anyons topological quantum computation}

\author{Mohamed Taha Rouabah}

\address{Laboratoire de Physique Math\'{e}matique et Subatomique, Fr\`{e}res Mentroui University Constantine 1, Ain El Bey Road, 25017, Constantine.}

\ead{rouabah.taha@umc.edu.dz}

\begin{abstract}
Topological quantum computation is an implementation of a quantum computer in a way that radically reduces decoherence. Topological qubits are encoded in the topological evolution of two-dimensional quasi-particles called anyons and universal set of quantum gates can be constructed by braiding these anyons yielding to a topologically protected circuit model. In the present study we remind the basics of this emerging quantum computation scheme and illustrate how a topological qubit built with three Fibonacci anyons might be adopted to achieve leakage free braiding gate by exchanging the anyons composing it.
A single-qubit braiding gate that approximates the Hadamard quantum gate to a certain accuracy is numerically implemented using a brute force search method. The algorithms utilized for that purpose are explained and the numerical programs are publicly shared  for reproduction and further use.
\end{abstract}

\section{Introduction}
Quantum computing promises exceptionally powerful computational capacities which might help make a giant leap in solving problems that are too complex for current \textit{classical} computers \cite{Benioff1980, Feynman1982, Nielsen2010, Arute2019, Gyongyosi2019}. However, one of the main obstacle to efficient quantum computation is decoherence. This latter quickly destroys the information in a superposition of states encoded in a quantum computer due to environment disturbances, thus seriously compromising long computations \cite{Schlosshauer2005, Schlosshauer2019}.
Many ways are being investigated in order to overcome this difficulty. Most of them are focusing on isolating the physical mediums of qubits from outside perturbations, along with finding ways to correct the errors caused by the fragility of such supports 
\cite{Calderbank1996,Chiaverini2005,Devitt2013}.
Beside these seemingly straightforward approaches to handle the decoherence problem, a radically different  method termed topological quantum computing (TQC) emerged from a particularly brilliant idea: instead of hiding the information from the environment, one can protect it by encoding it in a non-local property of the system, its topological state, thus concentrating efforts on making the qubits harder to alter without being  necessarily harder to reach. In other words, in TQC the information will be implemented in a more robust property of matter rather than fragile quantum physical systems (ion
traps, superconductors, photons, etc \ldots). The concept consists on using small collections of anyons  as qubits and performing computation with these qubits by  braiding the anyons, both within the collections that form the qubits and between different collections for multi-qubit gates.  Interestingly, a large class of anyon models have been shown to allow for universal quantum computation by braids \cite{Kitaev2003, Freedman2003,Collins2006, Pachos2012, Stanescu2017, Field2018, Belaloui2019}.

This work aims to answer--in the most accessible way possible--three fundamental questions in TQC: {\it why anyons ? How to implement a topological qubit with anyons ? How to manipulate a single qubit through  topological operations in order to compile a quantum gates ?}.
In section \ref{sec:qubit} we will first  present the two-dimensional quasi-particle called anyon which is the elementary constituent of this implementation. Its particular statistical properties are essential for implementing topological qubits and braiding gates in the TQC framework. 
Then, we will expose the Fibonacci anyon model which makes use of a specific class of anyons to first store information in their topological charge and then manipulate it by braiding the anyons. 
In section \ref{sec:gate}, we will explain how one can build a single-qubit braiding gate and present our approach for constructing braids that can approximate Hadamard quantum gate to a certain accuracy. We will also illustrate the algorithms we have used to obtain such an approximation. An ensemble of Python 3 programs have been tailored to implement those algorithms  and have been made publicly available on the repository \cite{Github_TQC} for reproduction and further use.

\section{Topological qubit}
\label{sec:qubit}

\subsection{Anyons}
\begin{figure}[t!]
    \centering
    \includegraphics[width=0.6\textwidth]{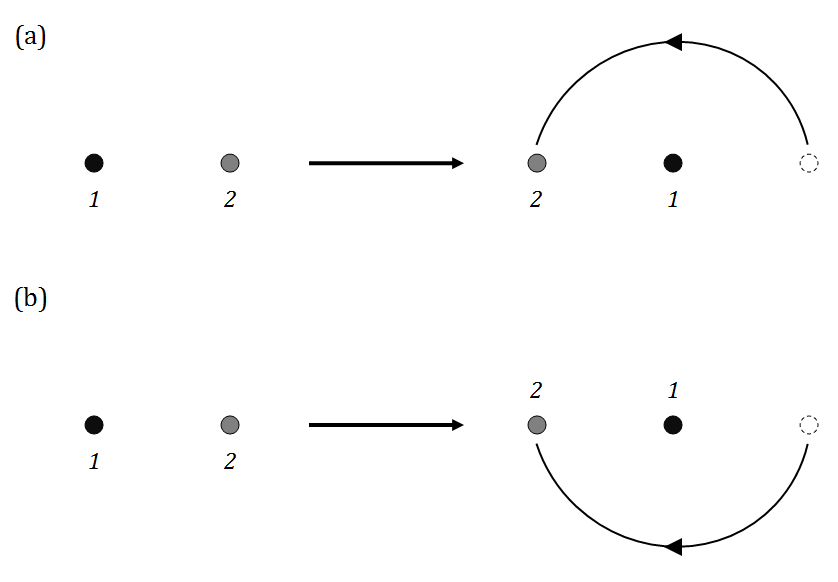}
    \caption{Exchanging two particles: the counterclockwise exchange (a) of particle 2 with respect to particle 1 corresponds to  $\Delta\phi = -\pi$ in Eq.\eqref{eq:phase} and the clockwise exchange (b) corresponds to $\Delta\phi = +\pi$ \cite{Lerda1992}.}
    \label{fig:1}
\end{figure}

Unlike in the usual three-dimensional (3D) quantum mechanics, where particles can either be fermions or bosons, corresponding respectively to half-integer and integer values of spin, two-dimensional (2D) particles can have a behavior which falls in neither of these two categories.
In quantum mechanics, the wave functions of two indistinguishable particles evolving in a 3D space may overlap, making it impossible to keep track of them separately. As a result, the description of a system made of such two particles requires the use of a common wave function for both of them.
Let $\psi(1,2)$ be such a wave function describing two identical particles: particle 1 and particle 2.
Rotating particle 2 around particle 1 by an angle $\Delta\phi$ gives rise to a complex phase such as
\begin{equation}\label{eq:phase}
\psi(1,2) \rightarrow \psi'(1,2) = e^{i \nu \Delta\phi} \psi(1,2),
\end{equation}
where $\nu$ is called the particle's statistics and the angle $\Delta\phi$ can be either $+\pi$ or $-\pi$, as shown in  Fig.\ref{fig:1}.
%
In 3D, both trajectories exposed in Fig.\ref{fig:1} are topologically equivalent--as one of them can always be continuously deformed into the other--giving the equality
\begin{equation}\label{eq:fig1}
e^{+i \nu \pi} = e^{-i \nu \pi},
\end{equation}
which is satisfied only for $\nu = 0, 1$.
That is, for the exchange described in Fig.\ref{fig:1}, Eq.\eqref{eq:phase} writes
\begin{equation}
\psi(2,1) = \pm \psi(1,2).
\end{equation}
The two possible phase factors resulting from  exchanging two indistinguishable particles in 3D correspond to the two only possible statistics namely those of bosons and fermions.

However, in 2D the two trajectories in Fig.\ref{fig:1} are not equivalent because there is no continuous deformation that transforms trajectory (a) to trajectory (b). Thus, no constraint equivalent to Eq.\eqref{eq:phase} applies on the statistics $\nu$ which should be arbitrary.
In this case, particles are not just either bosons or fermions; they can rather have \textit{any} statistics and are called \textit{anyons} \cite{Wilczek1982,Lerda1992,Khare2005}.
This is an important property for topological quantum computing. Indeed, anyons can show a richer statistical evolution when they are exchanged multiple times compared to fermions and bosons as the final phase factor will not be restricted to $\pm1$ but instead, it can take a wide range of values depending on the statistics $\nu$. 
Most importantly, given that anticlockwise exchange of two anyons is different from a clockwise exchange as described earlier, one can implement information in the topology of a system of multiple anyons and manipulate that information by \textit{braiding}  the world lines of these particles. We will go into more detail about this crucial property in section \ref{sec:qubit_implementation}.
%
%
%
It is worth to mention that no elementary particle is an anyon. Anyons can be created as quasi-particles in two dimensions. Experimentally, this can be achieved through the Fractional Quantum Hall Effect (FQHE)\cite{ Moore1991,de-Picciotto1998, Daijiro2002, Clarke2013}. The physics behind anyons creation and the FQHE is beyond the scope of this paper (see \cite{Tounsi2019} for a smooth introduction to quantum hall effect, anyons and topology).
Interestingly, anyons can be fused to produce different anyonic species forming a fusion space and can be created from another anyon or from the vacuum since each anyonic particle has an  anti-particle which is also an anyon. Anyonic species are distinguished by their conserved topological characteristics which we will call topological charges throughout this paper.  We will consider an anyon to have a charge of $1$, and the vacuum to have a charge of $0$. Thus, two anyons created from the vacuum will have a total charge of $0$, whereas two anyons which were the result of splitting an existing anyon will have a total charge of $1$. 
This topological charge is conserved inside a finite region of space  and will constitute the property in which information will be encoded to build a topological qubit. 

\subsection{Abelian and non-abelian anyons}
The reverse process of anyons pairs creation from either an anyon or the vacuum consists in the fusion of pairs of anyons to either an anyon or the vacuum, depending on how they were created. 
Simply speaking, classes of anyons which have only one fusion channel, i.e., that can fuse to the vacuum only or to an anyon only, are labeled abelian anyons whereas classes of anyons that can fuse either to form another anyon or to the vacuum are said to have multiple fusion channels and are called non-abelian anyons  \cite{Nayak2008}.
Different fusion channels of a system of many non-abelian anyons will correspond to degenerate wave functions for the involved system of anyons. 
These wave functions form the anyons' Hilbert space also referred to as fusion space.
Thereby, non-abelian anyons 
statistical evolution will be described by a unitary matrix phase factor whereas abelian anyons evolve with scalar phase factors and their fusion creates a trivial Hilbert space.
Consequently, the  expended Hilbert space of non-abelian anyons makes them a good candidate to perform quantum computation.


\subsection{Fibonacci anyon model}
Remind that in TQC, the system involves a collection of anyons described by an anyon model. The Hilbert space of the entire system is the fusion space of the anyons that the  basis consists of states characterized by the different ways in which all the anyons can fuse together.
For that reason, any TQC model should describe the way these anyons interact and how the system of anyons evolves over time. One can distinguish several anyon models depending mainly on the fusion rules that describes consistently all the  possible creation and fusion channels for a given set of anyon species.
%
Fibonacci model uses non-abelian anyons that can be obtained through the FQHE system at filling factor $12/5$ \cite{Nayak2008,Trebst2008}. 
%
If the process of fusion of two particles $a$ and $b$ into different possible resulting particles $c_i$ is written using the algebraic form
\begin{equation*}
a \times b = \sum_i N_{ab}^i c_i,
\end{equation*}
where $N_{ab}^i$ is the number of different possible processes of fusing $a$ and $b$ anyons to $c_i$ anyon,
we may detail the fusion rules of the Fibonacci model as follows
\begin{align} 
1 \times 0 &= 1,\\
0 \times 1 &= 1,\\
0 \times 0 &= 0,\\
1 \times 1 &= 1+ 0, \label{non-trivial}
\end{align}
with $N_{11}^1 = N_{11}^0 = 1$ where 0 refers to the vacuum and 1 refers to an anyon. In the Fibonacci model, we have only one anyonic species denoted by $1$ which is its own  anti-particle and the only non-trivial fusion rule is given by Eq.\eqref{non-trivial} in which  two Fibonacci anyons have two possible fusion processes. As mentioned above, in anyonic models every possible fusion process represents a quantum state. Hence, different fusion processes form a Hilbert space or a fusion space which dimension depends on the model and the number of anyons involved. Fibonacci anyons have fusion space of dimensions 1, 1, 2, 3, 5, 8, 13 \ldots bringing to mind the Fibonacci sequence  \cite{Trebst2008}.

\subsection{Implementing topological qubit in the Fibonacci model}
\label{sec:qubit_implementation}
\begin{figure}[t!]
    \centering
    \includegraphics[width=0.4\textwidth]{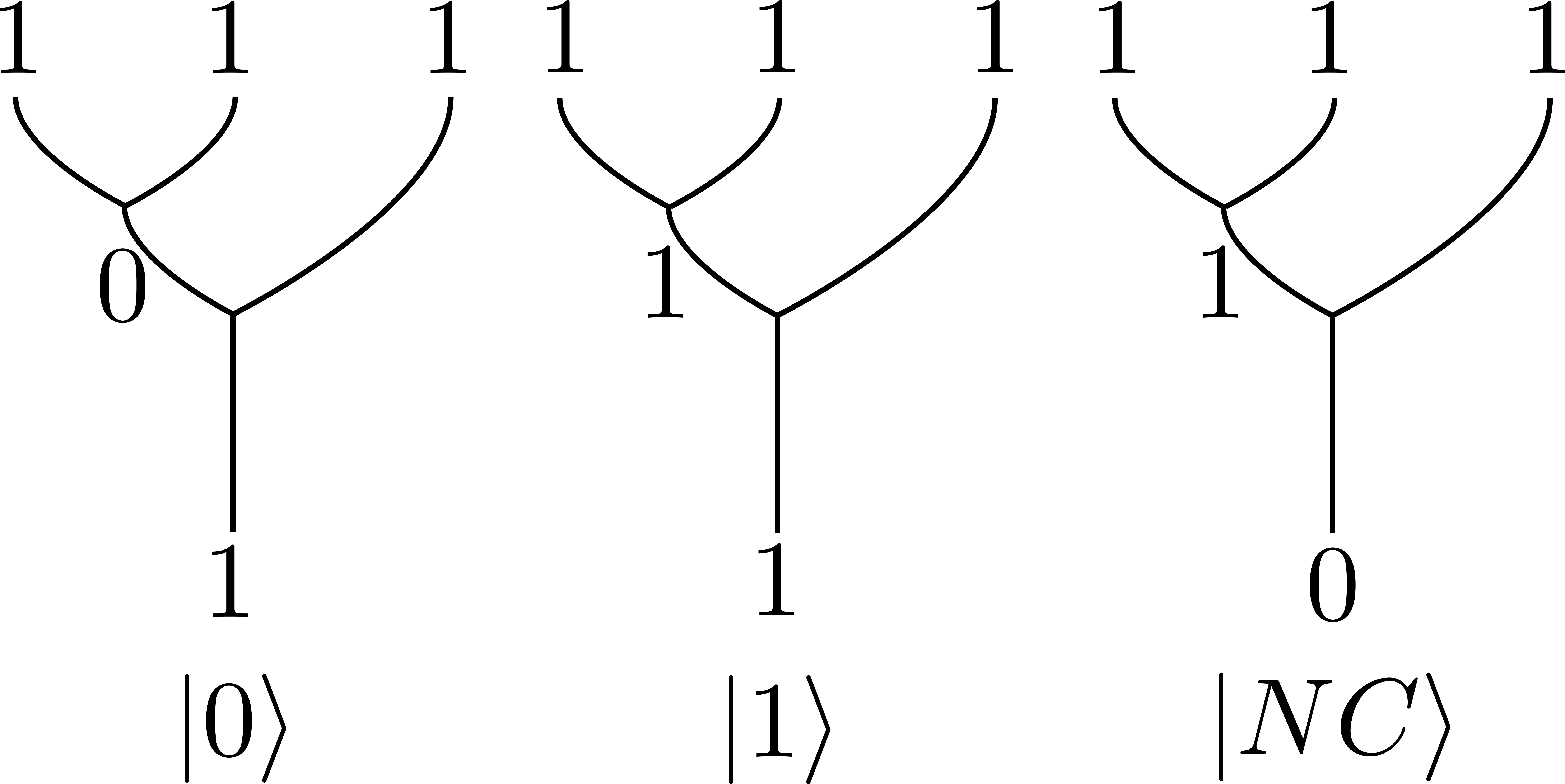}
       \caption{Qubit encoded in three Fibonacci anyons. Information is stored in the topological charge of anyons pairs and qubit states are represented in function of the succession of fusion results. From left to right we show fusion processes forming the qubit states $\ket{0}$, $\ket{1}$ and the non-computational state $\ket{NC}$. Given that the overall topological charge is conserved by braiding anyons from the same qubit, it is possible to omit the state $\ket{NC}$ by preparing the anyons in such a way that their final outcome will be $1$.}
    \label{fig:10}
\end{figure}

%
Assuming that it would be experimentally challenging to control multiple anyons, the optimal number of anyons from which to build a single qubit might be the minimal number of anyons that actually can provide a qubit. This would be two anyons, with the qubit of information stored in the topological charge, i.e., fusion channel of  two non-abelian anyons. However, with two-anyon qubits, no non-trivial gates can be achieved by braiding anyons within a single qubit.  
Qubits with three anyons solve this issue by implementing the information in the topological charge of the two leftmost or rightmost non-abelian anyons as will be illustrated latter in this section. This configuration can even allow for a computationally universal set of single-qubit operations that are performed by braiding only the anyons that form the qubit.
For instance, three Fibonacci anyons can be prepared to fuse in left-to-right order into a well defined sequence of intermediate and final anyons. From the fusion rules described previously, using three initial anyons, three possible fusions processes can be obtained corresponding to the three possible states of the three-anyon qubit as Fig.\ref{fig:10} shows.
This implementation is very practical, however, it leads to the apparition of a third state in addition to the  desired $|0\rangle$ and $|1\rangle$ qubit states which is the non-computation state that we will label $|NC\rangle$. 
This inadequation occurs when one aims to represent a quantum system supposed to evolve in a $d$-dimensional Hilbert space ($d=2$ for a qubit) using an anyonic system that have fusion space of larger dimension. As previously shown, three Fibonacci anyons with arbitrary total charge have fusion space of dimension 3. Consequently, if such anyons are used to represent a qubit the non-computational state $\ket{NC}$ will arise. This may suggest that when carrying operations on topological qubits, \textit{information leakage} may occur from the set of qubit states $\{ \ket{0}, \ket{1} \}$ to the third state $\ket{NC}$ when anyons are exchanged to perform braiding gates \cite{Xu2008,Ainsworth2011}. Fortunately, in single-qubit representation leakage errors are not an issue since the non-computational state is not correlated to the other qubit states.
In fact, the quantum number that stores the information being the topological charge of the two leftmost anyons of the qubit as shown in Fig.\ref{fig:10}, the computational Hilbert space is taken to be  a subspace of the full topological Hilbert space of the system corresponding to the tensor product of these two anyons fusion states.
Consequently,  any unitary operation acting on the three-anyon qubit in the total Hilbert space can be reduced to a ($2\times 2$) representation acting on the computational Hilbert space spanned by the basis $\{ \ket{0}, \ket{1} \}$ and returning  a linear composition of these two states only, thus allowing to construct leakage free single-qubit braiding gates. It is worth noting that in two-anyon qubits, leakage enters even at the level of single-qubit gates. Indeed, using two-anyon qubits approximating single-qubit gates will require special braidings of anyons that are part of different qubits. See \cite{Ainsworth2011} for a study that examines under which circumstances one may have models where all one-qubit and/or all multi-qubit braiding gates are entirely leakage free.
Noteworthy, contrary to traditional qubit implementations and thanks to its distinctive topological properties,  anyon-based topological qubit is resilient to small external perturbations yielding to a fault-tolerant quantum computation. The robustness of this model lies on the fact that moving anyons around each other in an undesirable way is most unlikely when these particles are kept far apart \cite{Hormozi2009}.
%

\subsection{Manipulating anyons in the 
Fibonacci model}
Once topological qubit implemented, one can exchange the anyons in order to modify the intermediate outcome of their fusions and then manipulate the quantum state of the qubit.
In what follows we will present the $F$ and $R$ moves based on which the elementary braiding operations of the model will be defined. 
Those latter applied on the topological qubit will allow to handle its quantum state within the computational Hilbert space in order to achieve information processing.
\subsubsection{Changing the order of fusion: the F matrix \ }
\begin{figure}[t!]
    \centering
    \includegraphics[width=0.35\textwidth]{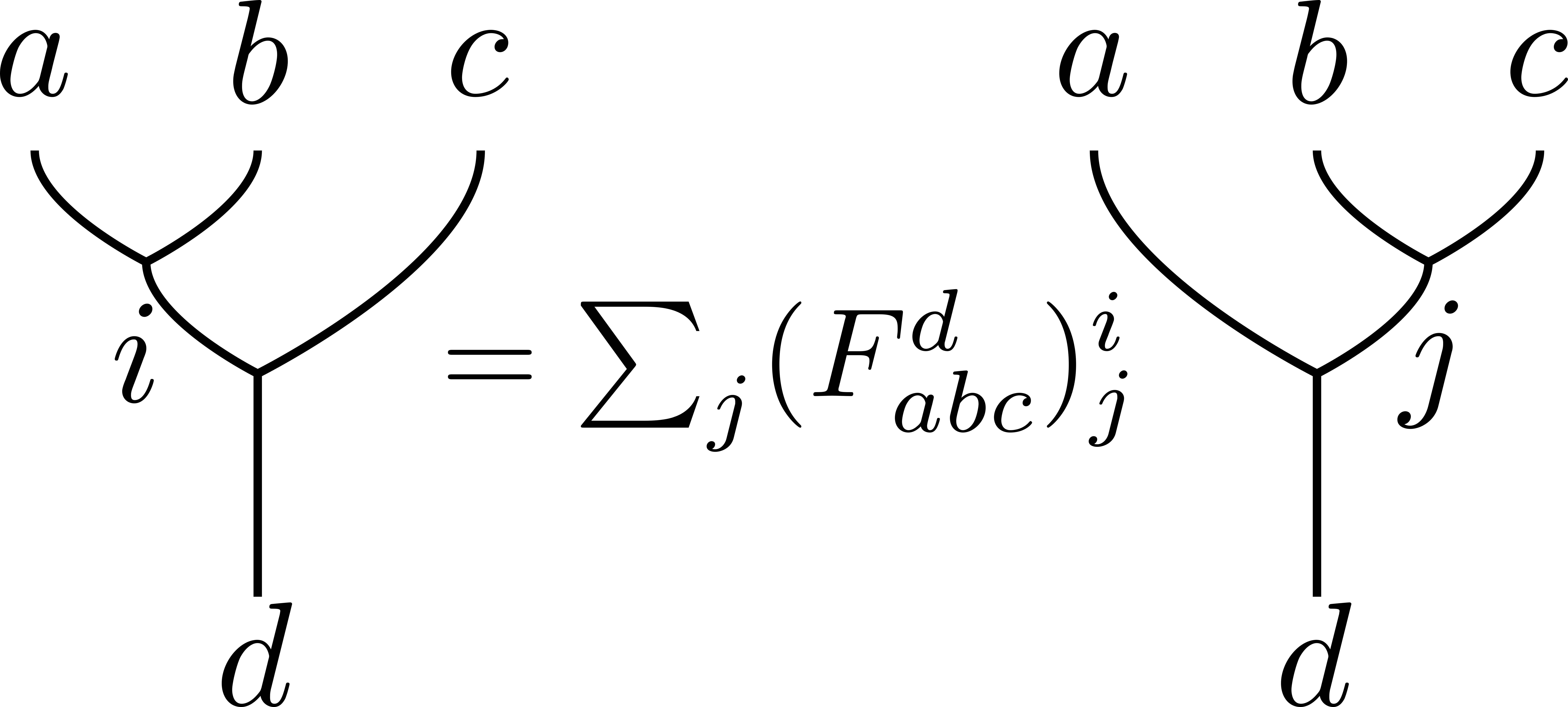}
    \caption{The two possible orders of fusions of the initial anyons $a,b$ and $c$ are mapped to each other using the $F$ matrix. Anyons $i$ and $j$ are the intermediate  results of the first fusion and $d$ is the final anyon.}
    \label{fig:4}
\end{figure}

Provided three anyons or more, one can either fuse them starting by fusing the leftmost pair first or the rightmost pair first. Notice that the final result will not be affected since the total charge of these anyons before the fusion is  conserved.
Given three anyons $a, b$ and $c$ to be fused to an anyon $d$, mapping one order of fusion to another is done via one application--or multiple applications for more than three anyons--of the $F$ matrix as shown in Fig.\ref{fig:4}.
To compute the elements  $(F^{d}_{abc})^i_j$  a consistency equation is constructed as follows:
consider four anyons labeled 1, 2, 3, and 4 from left to right and a fusion process in which the leftmost anyon is fused in a left-to-right fusion process until we end up with only one anyon. The right-to-left fusion will end up with the same result through different intermediate fusions (see Fig.\ref{fig:5}).  Since these two trajectories are equivalent, it is possible to write
\begin{equation} \label{pentagoneqt}
(F^{e}_{jcd})^i_k (F^{e}_{abk})^j_m  = \sum_l (F^{i}_{abc})^j_l (F^{e}_{ald})^i_m (F^{m}_{bcd})^l_k.
\end{equation}
This is called the \textit{pentagon identity}, in reference to the geometrical pattern in Fig.\ref{fig:5} where the left-hand side of the above equation corresponds to the upper path whereas the right-hand side corresponds to the lower path.
Forasmuch as $(F^{d}_{abc})^i_j$ has six indices, each of which taking either the value 0 for the vacuum or 1 for an anyon, there are 16 unitary matrices $F_{abc}^d$ following the $2^4$ possibilities of the set $\{a,b,c,d\}$ with 4 elements each depending on the $2^2$ values of the set $\{i,j\}$.
Using Eq.\eqref{pentagoneqt} with the fusion rules of the Fibonacci model, all the elements  $(F^{d}_{abc})^i_j$ in the model can be calculated.
For example, the $F$ matrix mapping the two possible fusion orders of three anyons (as in Fig.\ref{fig:4}) in the Fibonacci model for $a, b, c, d = 1$ is written in the basis \{$\ket{1\times 1\rightarrow 0}, \ket{1\times 1\rightarrow 1 }$\}) as follow:
\begin{align} 
\label{F1111}
F^{1}_{111} = 
    \begin{pmatrix}
		(F^{1}_{111})^0_0 & (F^{1}_{111})^0_1 \\
	(F^{1}_{111})^1_0 & (F^{1}_{111})^1_1
	\end{pmatrix}
=   \begin{pmatrix}
		\tau & {\sqrt{\tau}} \\
		{\sqrt{\tau}} 
        & -{\tau}
	\end{pmatrix}, 
\end{align}
where 
$\tau = (\sqrt{5}-1)/2 \approx 0.618$ is the inverse of the golden ratio.
Notice that the ($2\times2$) representation in Eq.\eqref{F1111} consists of a reduced representation of $F_{111}^1$ acting on the three  anyons computational Hilbert space only  \cite{Hormozi2007}.
Also, from a quantum computational point of view, it is worthy to notice that the matrix  $F_{111}^1$  might serve to change the fusion basis by just changing the order of fusion such that:
\begin{align}
\ket{0'}&=(F_{111}^1)_0^0 \ket{0} + (F_{111}^1)_1^0 \ket{1},\\
\ket{1'}&=(F_{111}^1)_0^1 \ket{0} + (F_{111}^1)_1^1 \ket{1} ,
\end{align}
where primed states represent right-to-left ordered fusion states and unprimed states represent left-to-right ordered fusion states. 
\begin{figure}[t!]
    \centering
    \includegraphics[width=0.28\textwidth]{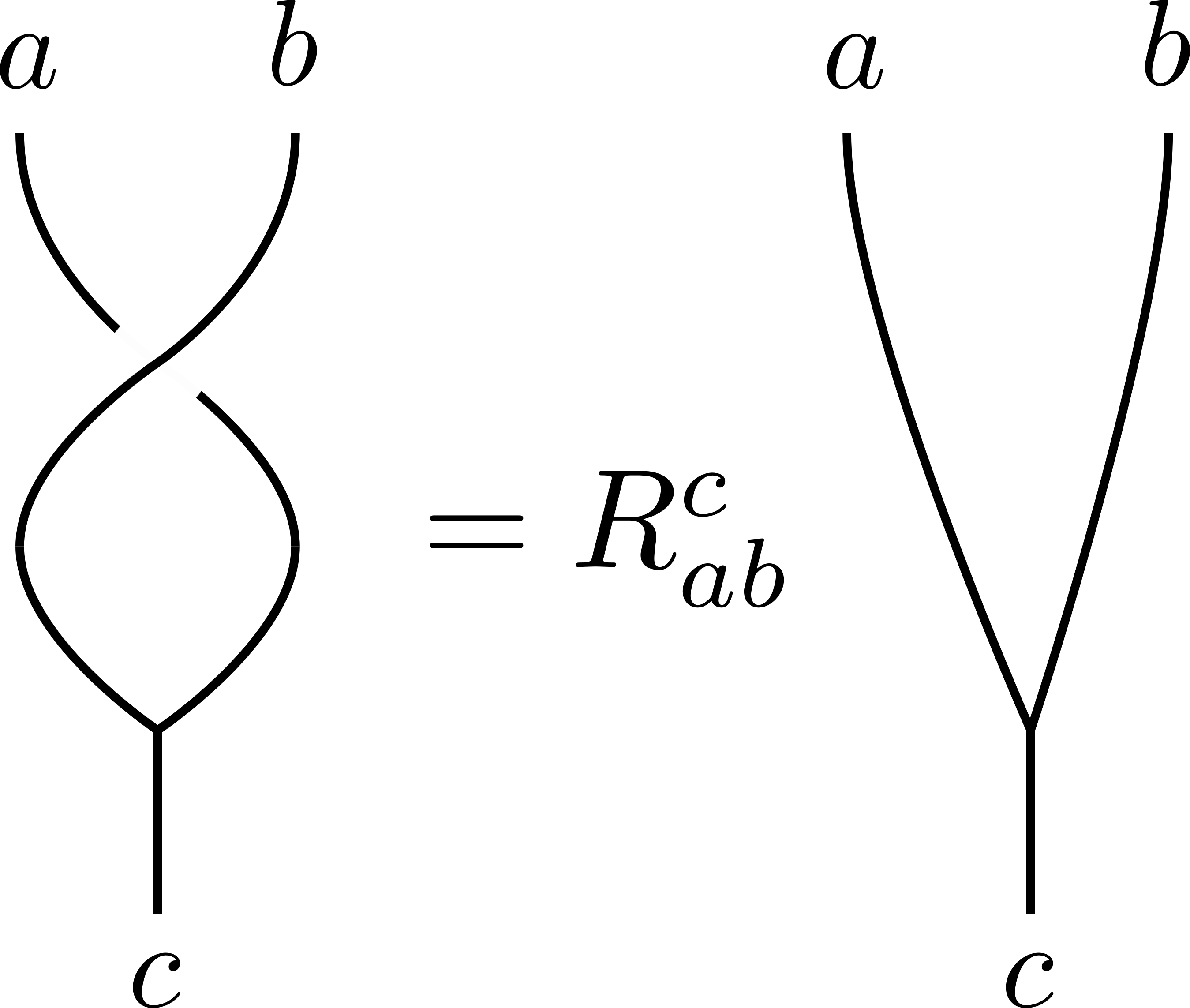}
    \caption{The $R$ matrix represents the operation corresponding to a clockwise interchanging of a and b.}
    \label{fig:6}
\end{figure}

\subsubsection{Interchanging anyons: the $R$ matrix \  }
\begin{figure}[t!]
\begin{subfigure}{1\textwidth}
    \centering
    \includegraphics[width=0.7\textwidth]{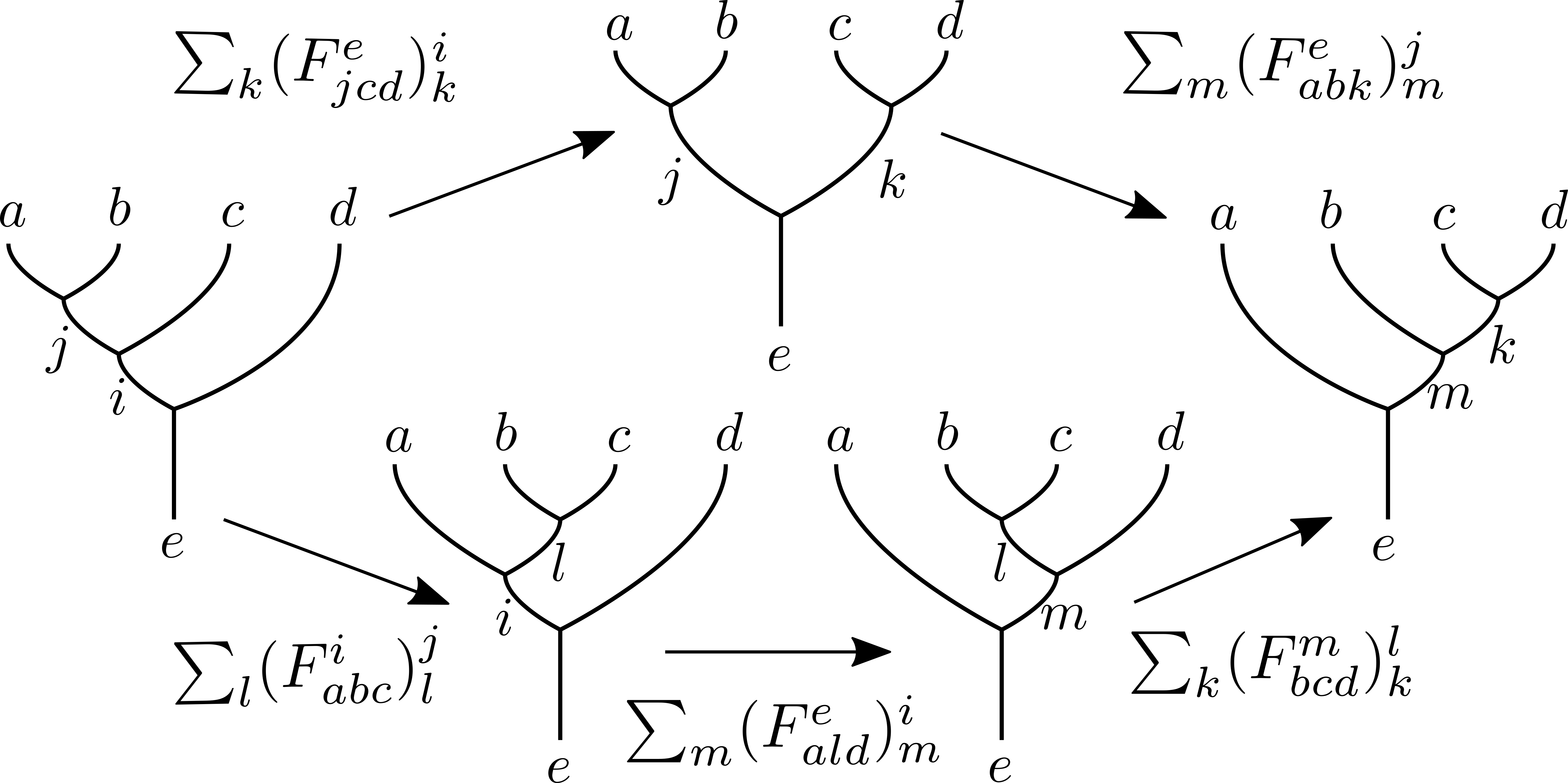}
    \caption{The two series of $F$ moves linking the left-to-right  to the right-to-left fusion processes of four anyons form a pentagon diagram. The equality between the two series constitute the pentagon identity \eqref{pentagoneqt}.}
    \label{fig:5}
\end{subfigure}
\begin{subfigure}{1\textwidth}
    \centering
    \includegraphics[width=0.7\textwidth]{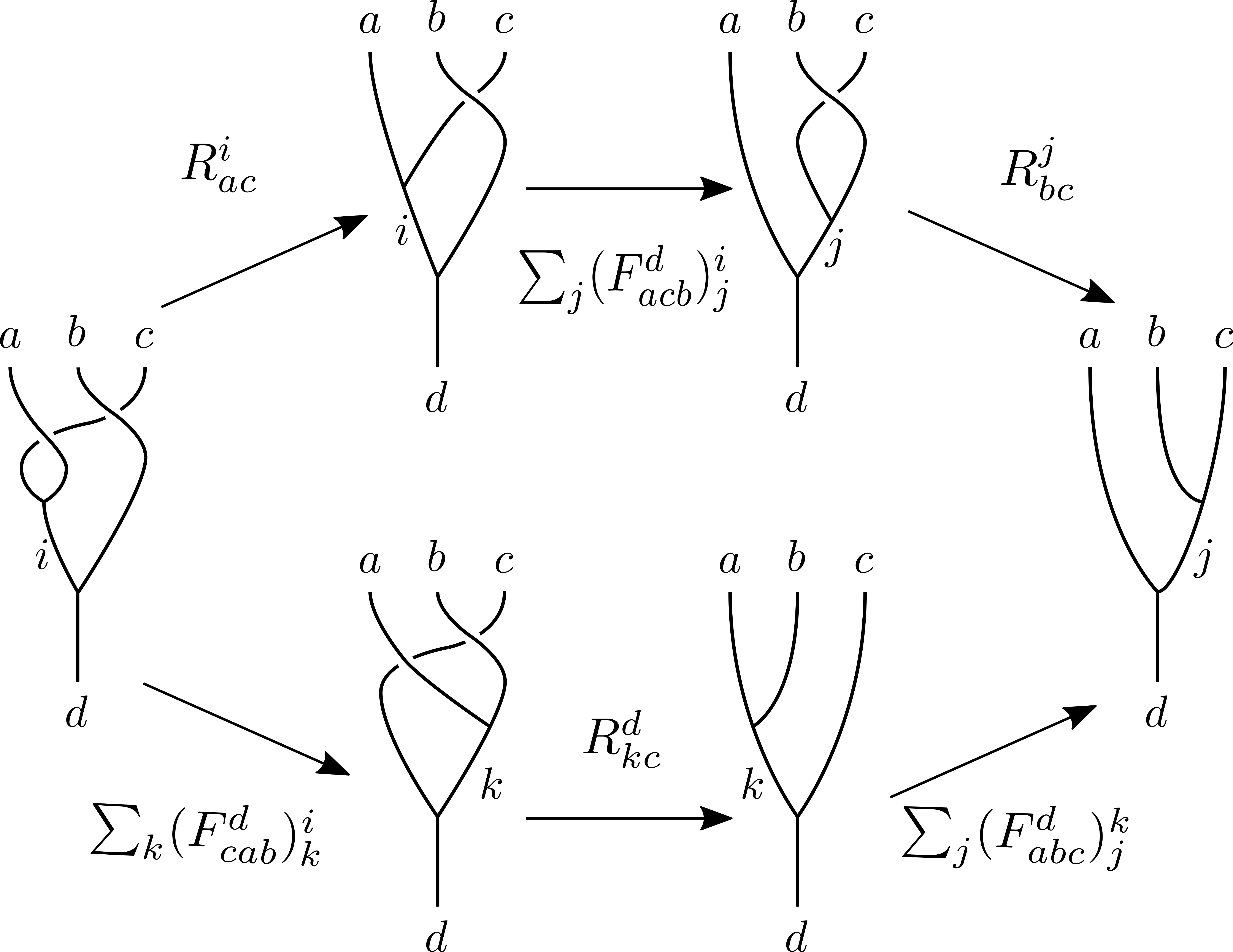}
    \caption{
    Hexagonal shape displaying the two possible $R$ and $F$ moves series mapping the leftmost state to the rightmost state  over the fusion space of three anyons. The identity between those two series constitute the consistency equation \eqref{hexagoneqt}.
    }
    \label{fig:7}
\end{subfigure}
    \caption{Pentagon and hexagon diagrams.}
\end{figure}

The other way to manipulate anyons is to interchange them clockwise or anti-clockwise before fusion. Interchanging two anyons will give rise to a phase factor.
The matrix $R_{ab}^c$ corresponds to the operation of interchanging two anyons $a$ and $b$ clockwise before fusing them into an anyon $c$. The inverse matrix corresponds to the anti-clockwise interchange.
Similarly to the $F$ matrix, a consistency equation can be constructed for the $R$ matrix based on the equivalence between two possible ways of application of $F$ and $R$ moves to link the left-to-right and the right-to-left fusion--corresponding to the two different fusion states--of three anyons. This equation known as  the \textit{hexagon identity} is composed of both the $F$ and $R$ matrices as follow (see Fig.\ref{fig:7}):
\begin{equation} 
\label{hexagoneqt}
\sum_k (F^{d}_{cab})^i_k  R^{d}_{kc} (F^{d}_{abc})^k_j  =  R^{i}_{ac} (F^{d}_{acb})^i_j  R^{j}_{bc}.
\end{equation}
It is clear that in order to compute the elements of the $R$ matrices using the hexagon identity, one has to compute first the $F$ matrices.
The matrix $R_{11}$ needed to interchange two anyons (corresponding to two topological charges of 1) in the Fibonacci model is given by 
\cite{Hormozi2007} under the form 
\begin{equation} \label{R11}
R_{11} =  \begin{pmatrix}
		e^{-i4\pi/5} & 0 \\
		0 & e^{i3\pi/5}
	\end{pmatrix}.
\end{equation}
%
%

\subsubsection{Elementary braiding in the Fibonacci model \ }
Since interchanging anyons clockwise and anticlockwise are not equivalent operations, the world lines of these anyons are said to be braided after a sequence of exchanges \cite{Bonesteel2005}.
The clockwise elementary braiding operator of the two leftmost pair in a set of multiple anyons, e.g. the pair \{$a,b$\} in Fig.\ref{fig:4}, is referred to as $\sigma_1$, and the anticlockwise elementary braiding operator of the same pair will naturally be its inverse $\sigma_1^{-1}$. They are simply equivalent to an $R$ and $R^{-1}$ moves respectively. 
The reduced representation of the elementary braiding operator $\sigma_1$ for a system of three anyons is given by \cite{Hormozi2007}:
\begin{equation} \label{sigma1}
\sigma_1 = \begin{pmatrix}
		e^{-i4\pi/5} & 0 \\
		0 & e^{i3\pi/5} 
	\end{pmatrix},
\end{equation}
%

%
The braiding of the next pair  of anyons, e.g. \{$b,c$\} in Fig.\ref{fig:4}, will be achieved through the operator $\sigma_2$. This latter is obtained from the successive applications of the matrices $F$, $\sigma_1$, and  $F^{-1}$:
\begin{equation}
\sigma_2 = F^{-1} \sigma_1 F.
\end{equation}
Following \cite{Hormozi2007} the operator $\sigma_2$ for a system of three anyons will be:
\begin{equation} \label{sigma2}
\sigma_2 = \begin{pmatrix}
		-\tau e^{-i\pi/5} & -\sqrt{\tau}e^{-i3\pi/5}\\
		\sqrt{\tau}e^{-i3\pi/5} & -\tau  \\
	\end{pmatrix},
\end{equation}
All possible world lines braids of the three anyons composing the topological qubit we implemented earlier can be obtained by multiplying elementary braiding operators $\sigma_i$  and their inverses. The braiding operators form the \textit{braiding group} $B_n$ which governs the exchanges of $n$ anyons with $n-1$ generators. The group generators satisfy the relations:
\begin{align}
\sigma_i\sigma_j &= \sigma_j\sigma_i, \hspace{1.17cm} \text{ for } \ |i-j|\ge 2, \nonumber\\
\sigma_i\sigma_{i+1}\sigma_i &= \sigma_{i+1}\sigma_i\sigma_{i+1}, \ \text{ for } \ 1 \le i \le n,\\
\sigma_i{\sigma_i}^{-1} &= {\sigma_i}^{-1}\sigma_i = e, \nonumber
\end{align}
where $e$ is the identity element of the group \cite{Khiat2019}.

\section{Topological quantum gate}
\label{sec:gate}
\begin{figure}[t!]
    \centering
    \includegraphics[width=0.46\textwidth,height=2.2cm]{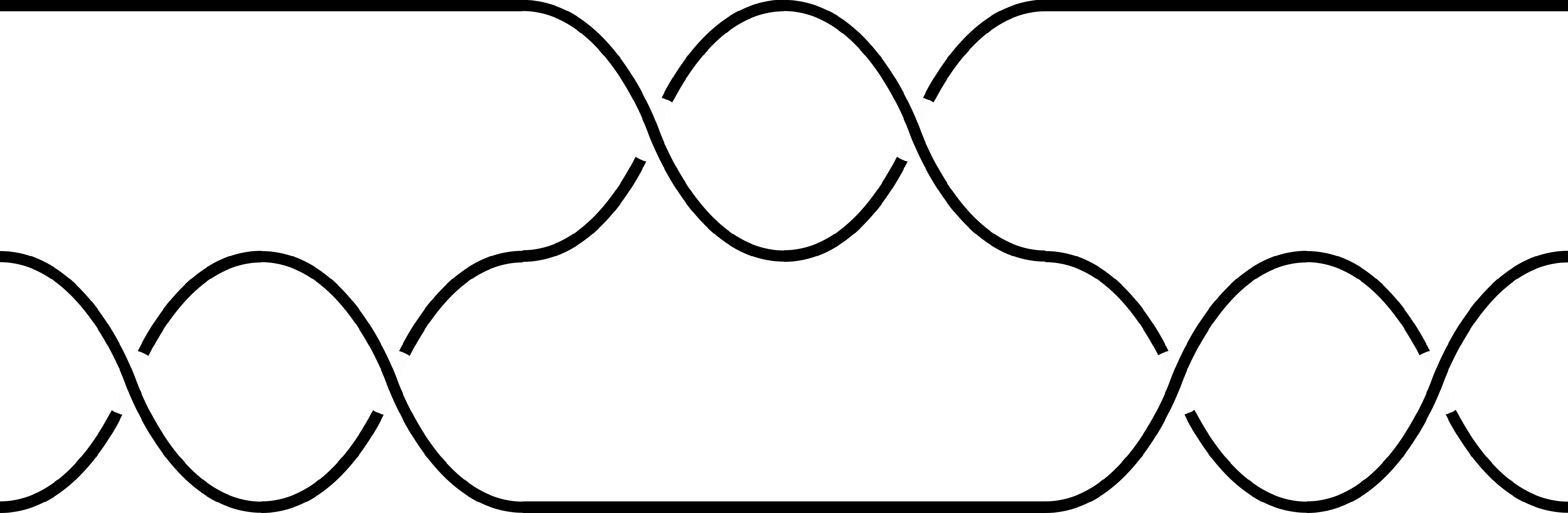}
    \caption{The braiding pattern  representing the quantum gate $G = \sigma_1^{-2} \sigma_2^2 \sigma_{1}^{2}$. Time points to the right.}
    \label{fig:11}
\end{figure}

Applying a quantum gate on a topological qubit consists in \textit{braiding} its constituent anyons. While braiding anyons from the same qubit does not change the overall topological charge of the qubit, braiding anyons from different qubits around each other may change the total topological charge of the qubits and this will result in part of the information \textit{leaking} out of the computational Hilbert space \cite{Ainsworth2011}. In this section we will focus on single-qubit braiding gate for a leakage free quantum computation.
Mathematically speaking, quantum evolution of the topological qubit is achieved by successive applications of the elementary exchanges  $\sigma_1$ and $\sigma_2$ on its anyons pairs leading to a methodical manipulation of its quantum state in the computational Hilbert space. 
As an example, Fig.\ref{fig:11} shows the braid representing the braiding gate operator $G$ defined as
\begin{equation} \label{G}
G = \sigma_1^{-2} \sigma_2^2 \sigma_{1}^{2}.
\end{equation}
Notice that the braiding representation   in Fig.\ref{fig:11} is expressed in inverse order compared to the braiding operator $G$ due to the fact that, mathematically, the elementary braiding operator $\sigma_i$ corresponding to the first braid undergone by one pair of anyons (leftmost on  Fig.\ref{fig:11}) should be applied first on the qubit state (rightmost on the operator expression).
\subsection{The Solovay-Kitaev theorem}
At this point, important question raises: {\it 
Is it possible to construct an equivalent braid for any quantum gate ?} The answer is yes and it comes from the so-called Solovay-Kitaev theorem stating the following  \cite{Nielsen2010}:\\

\textit{Let $G$ be a finite set of elements in $SU(d)$ containing its own inverses, such that the image of $G$ is dense in SU(d), and let a desired accuracy $\epsilon$ be given. There exists a constant $c$ such that for any $U \in SU(d)$ there exists a finite sequence $S$ of gates in $G$ of length $O(log^c(1/\epsilon))$ such that $d(U,S) < \epsilon$}.\\

Simply speaking, this theorem states that it is always possible to approximate any unitary quantum gate $U$ to a desired accuracy $\epsilon$ using a long enough braid $S$. The length of a braid $S$ being defined as the number of elementary braiding operators $\sigma_i$ it contains. Therefore, a topological quantum computer can be effectively described by the quantum circuit model provided that the elementary exchanges of the anyons in the qubits construct a universal set of gates generating all possible unitary quantum operations.

\subsection{From braids to weaves: simplifying the problem} \label{weaves}
Simon {\it et al.} showed in  \cite{Simon2006} that \textit{weaves} instead of braids are sufficient to construct any quantum gate. Mainly, rather than moving all the anyons of the qubit to make a braid, one can make a weave instead by only moving one anyon around the other anyons which will be kept fixed as shown on Fig.\ref{fig:11}. This implies that in a weave, only  pair powered elementary braiding operators will be used. Namely $\sigma_i^2$, $\sigma_i^4$, $\ldots$, and their inverses. 
Moreover, it is interesting to notice that the successive application of the $\sigma_i$ matrices \eqref{sigma1} and \eqref{sigma2} leads to a periodicity of the form:
\begin{equation}
\sigma_i^{10} = \sigma_i.
\end{equation}
Therefore,
\begin{align}
\sigma_i^{8} &= \sigma_i^{-2},\\
\sigma_i^{6} &= \sigma_i^{-4},\\
\sigma_i^{4} &= \sigma_i^{-6},\\
\sigma_i^{2} &= \sigma_i^{-8}.
\end{align}
That is, any weave can be expressed in term of the operators $\sigma_i^{2}$, $\sigma_i^{4}$  and their inverses only.

\subsection{Approximating a single-qubit quantum gate}\label{sec:singlegates}
In this section we will present a numerical brute force search method to achieve the appropriate series of weaves with a given length which approximate the best a target quantum gate.
As already mentioned, applying a quantum gate to a single qubit in the Fibonacci model can be accomplished by braiding (weaving) the three anyons of a topological qubit. 
To find the braid that corresponds to a given quantum gate we numerically perform a brute force search where we try all the possible braiding patterns up to a given braid length (number of exchanges), and keep the one which approximates the best the target gate at the end.
In order to estimate how close a produced braid is to the target quantum gate we will use, for the sake of consistency with literature, the following global phase invariant error metric originally introduced by Bonesteel {\it et al} in \cite{Bonesteel2005} and used in multiple studies including the recent extensive work by Field \textit{et al.} \cite{Field2018}: 
\begin{equation} \label{eq:bonesteel}
\varepsilon(W,U) =  \sqrt{2 - |tr(W U^\dagger)|},
\end{equation}
where $W$ is the operator obtained by braiding the three anyons, $U$ is the target quantum gate and $\varepsilon$ is the error of the braid indicating the accuracy of the brute force search operation. 

At this point, we have all the necessary tools to take on the main task of this section: approximating a target single-qubit quantum gate by weaving the anyons of a three-anyon topological qubit. 
One of the most widely used single-qubit quantum gate is the Hadamard gate. We will take it as an example in this section and we will present the algorithms we use for approximating that elementary quantum gate by weaving anyons.
We first remind that the Hadamard gate $H$  is a quantum gate which acts on a single-qubit as follows:
\begin{align}
H \ket{0} &= \frac{1}{\sqrt{2}}(\ket{0} + \ket{1}),\\
H \ket{1} &= \frac{1}{\sqrt{2}}(\ket{0} - \ket{1}).
\end{align}
Using the representation
\begin{equation}
\ket{0} = \myvector{1 \\ 0}, \hspace{1cm} \ket{1} = \myvector{0 \\ 1},
\end{equation}
the matrix form of the Hadamard  gate is then given by
\begin{align}
H &= \frac{1}{\sqrt{2}} \myvector{1 & 1 \\ 1 & -1}.
\end{align}
Let us now expose the required steps to construct the braid operator representing the approximate braiding equivalent of the Hadamard gate.

\subsubsection{Step 1: Generating the set of possible weaves}
\begin{figure}[t!]
    \begin{subfigure}{0.5\textwidth}
\centering
\includegraphics[width=0.8\linewidth,height=2.5cm]{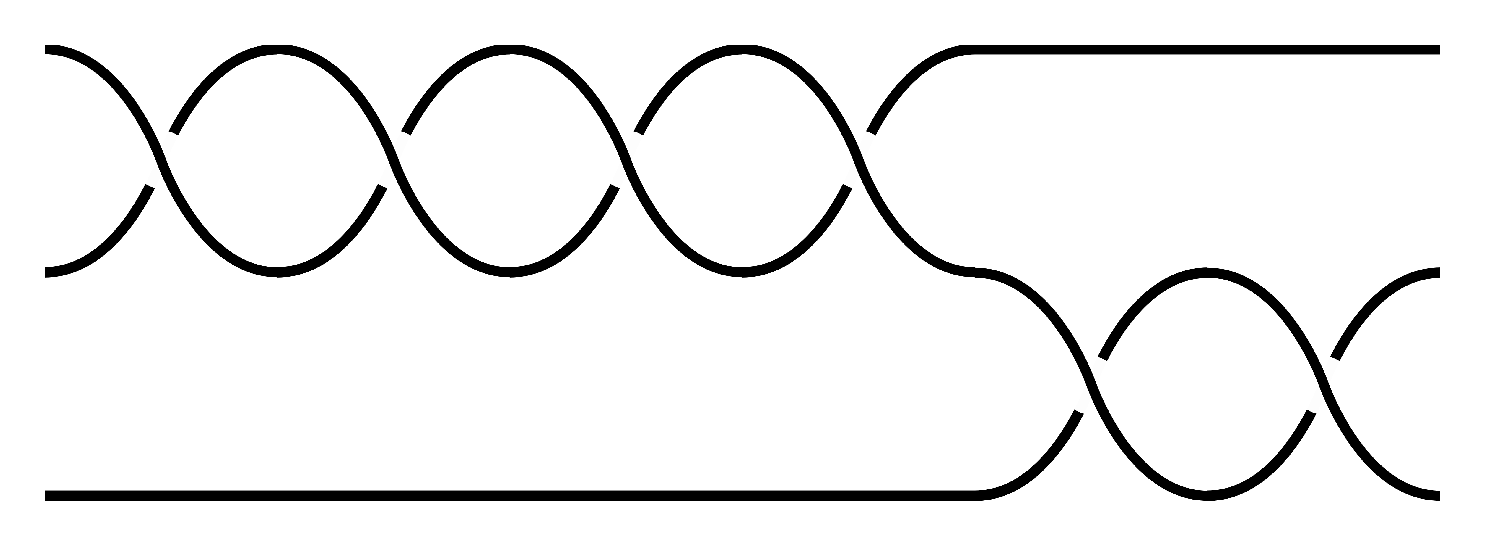} 
\label{fig:12a}
    \end{subfigure}
    \begin{subfigure}{0.5\textwidth}
\centering
\includegraphics[width=0.8\linewidth,height=2.5cm]{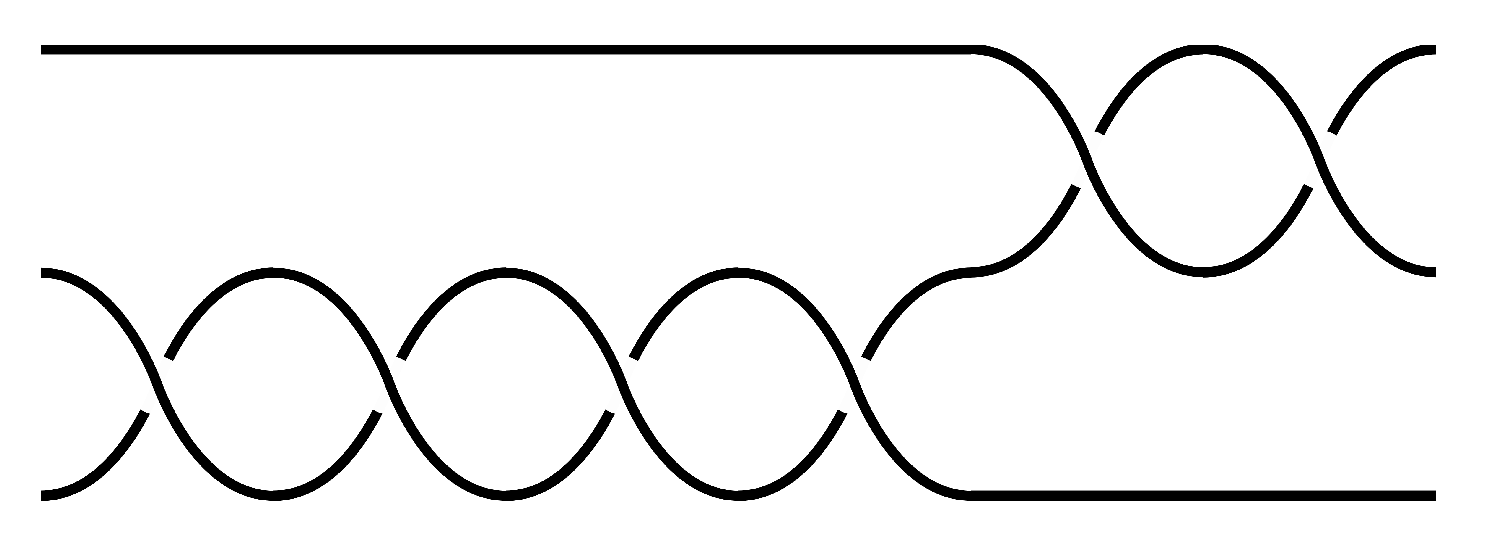}
\label{fig:12b}
    \end{subfigure}
\caption{The two possible weaves representations for the sequence $W_6$ in Eq.\eqref{eq:W6} corresponding to the weaving operators $\sigma_1^{+2} \sigma_2^{+4}$ (left) and  $\sigma_2^{+2} \sigma_1^{+4}$ (right). Time points to the right. Remind that the rightmost element in the weaving operator  corresponds to the leftmost move on the weave representation.}
\label{fig:12}
\end{figure}

Since we will perform a brute force search to look for the best weave approximating the Hadamard gate, we first need to generate \textit{all} the possible weaves up to a given length, i.e., number of $\sigma_i$ operators composing the weave.
If we consider for instance a maximum braid length $L = 30$, we will need to generate all the possible weaves with $L$ elementary braiding operators taking into consideration the periodicity exposed in Sec.\eqref{weaves}. Concretely, generating the weaves consists in generating the sequence of powers for the successive braiding operators $\sigma_i$. Since in a weave those powers are pair numbers with a minimum absolute value of 2, the maximum length possible for the powers sequence in order to construct a braid of length $L$ is $L/2$.
Thereby, a 30 elementary moves weave will be represented by the set $W_k$ ($k = 0, 1, 2, \ldots$) containing a sequence of the 15 powers of the braiding operators composing that  weave.
The very first weave is the trivial one  approximating the identity gate. Its powers are all zeros:
\begin{equation}
W_0 = \{0,0,0,0,0,0,0,0,0,0,0,0,0,0,0\}.
\end{equation}
We then start counting from the right using the basis $\{-4, -2, +2, +4\}$ which corresponds to the only needed powers of $\sigma_i$ in a weave due to the periodicity we discussed earlier. The power coming after $0$ being $+2$, the next weaves will then have the forms:
\begin{align} 
W_1 &= \{0,0,0,0,0,0,0,0,0,0,0,0,0,0,+2\},\\
W_2 &= \{0,0,0,0,0,0,0,0,0,0,0,0,0,0,+4\},\\
W_3 &= \{0,0,0,0,0,0,0,0,0,0,0,0,0,+2,-4\},\\
W_4 &= \{0,0,0,0,0,0,0,0,0,0,0,0,0,+2,-2\},\\
W_5 &= \{0,0,0,0,0,0,0,0,0,0,0,0,0,+2,+2\},\\
W_6 &= \{0,0,0,0,0,0,0,0,0,0,0,0,0,+2,+4\},
\label{eq:W6}\\
\vdots \ \ &{} \hspace{3cm} \vdots \nonumber
\end{align} 
and so on until all the $4^{L/2}$ elements $W_k$ are built.
%
%
Notice that we constructed the sequences $W_k$ that contain the powers of the braiding operators composing the weaves but we have not stated whether we should start those weaves operators with $\sigma_1$ or $\sigma_2$. Actually, both cases must be considered in order to completely cover the set of all possible weaves. That is, every set $W_k$ corresponds to two different weaves operators having the same powers sequence but starting with different elementary braiding operators. For instance, $W_6$ in Eq.\eqref{eq:W6} represents both $\sigma_1^{+2} \sigma_2^{+4}$ and $\sigma_2^{+2} \sigma_1^{+4}$. Those two weaves are shown in Fig.\ref{fig:12}.

\subsubsection{Step 2: Computing the set of all possible weaves}
After generating all the possible weaves of length $L = 30$ or less, we must compute the corresponding matrices in order to compare them with the Hadamard gate.
The process is straight forward: the matrix representation of every weave is obtained by multiplying the braiding operators $\sigma_i$ to the corresponding powers sequence  $W_k$ starting the weave with $\sigma_1$, then do the same starting with $\sigma_2$.

\subsubsection{Step 3: Searching for the best weave }
Now that we have generated the set of all possible weaves of length $L$, and computed the corresponding matrices we need to estimate the error in each weave  using the metric defined in Eq.\eqref{eq:bonesteel}. The braiding counterpart of the Hadamard quantum gate will then be represented by the weave that achieves the smallest $\varepsilon$.

\section{Results and Discussion}
\begin{figure}[t!]
    \centering
    \includegraphics[width=0.99\textwidth, height=2.6cm]{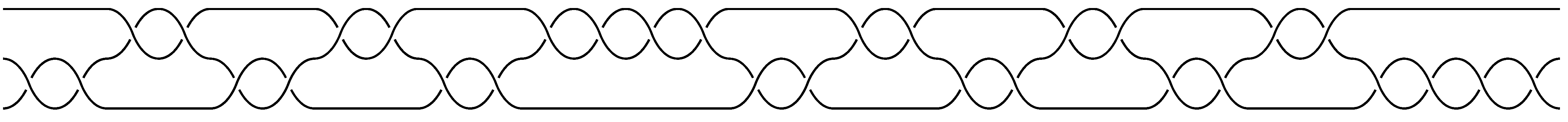}
    \caption{Wave representation of the operator $W_H$ constructed through the $30$ consecutive  braiding of  Eq.\eqref{eq:W_H_1}  approximating the Hadamard gate with an error of 0.00657. Time points to the right.}   \label{fig:13}
\end{figure}

Using tailored Python 3 programs, we have computed weaves which approximate the Hadamard gate up to a phase factor ($-i$), with length $L = 30$. The best weave determined with numerical brute force search corresponds to the braiding operator:
%
\begin{equation}
\label{eq:W_H_1}
   W_H = 
\sigma_1^{+4} \sigma_2^{-2} \sigma_1^{+2} \sigma_2^{-2} \sigma_1^{+2} \sigma_2^{+2} \sigma_1^{-2} \sigma_2^{+4} \sigma_1^{+2} \sigma_2^{-2} \sigma_1^{-2} \sigma_2^{+2} \sigma_1^{+2}, 
\end{equation}
which has the following matrix representation 
\begin{equation}
\label{eq:W_H_2}
W_H = -\frac{i}{\sqrt{2}} \myvector{1.0040+0.0056i & 0.9959-0.0048i \\ 0.9959+0.0048i & -1.0040+0.0056i}.
\end{equation}
In Fig.\ref{fig:13} we show the world lines evolution of the three anyons composing the topological qubit being braided to construct the operator $W_H$. This latter is a braiding  approximation of the Hadamard gate with the error
\begin{equation}
\varepsilon(W_H,H) = 0.00657 .
\end{equation}
Notice that the error in our approximation is of the same order as the approximation exposed in reference \cite{Field2018} with an error of 0.003 using 34 elementary braiding operation.
The programs we employed to implement this approximation are publicly available on the link provided in \cite{Github_TQC}. Our limitation to 30 braid long approximation is purely related to limited computational capacities. Possibilities to obtain a lower error through the adaptation of our programs to HPC deployment are under investigation.
%

Furthermore, as we seek for more accuracy with considering longer  braids, the problem of searching for weaves becomes harder to the point of being practically impossible due to very large computation time required. The bidirectional search is a way to optimize the brute force search using the same computational capacities. 
In this method, instead of just starting from the trivial identity weave and going up to the maximum length $L$, we also start the search in reverse: from the target gate towards the identity.
Thus, instead of generating all the possible weaves of length $L$ we will generate all possible half-weaves of length $L/2$ in one direction and the other separately.
At the end, we look for the half-weaves from different directions which meet at ``the middle" to compose the complete weaves of length $L$. From all possible complete weaves we will pick-up the one that approximates the best the target gate.
The only issue with this method is that it does not always give the best weave, as the bidirectional search only performs on a subset of all the possible weaves of maximum length $L$ composed of two possible half-weaves of length $L/2$ or less that meet on their extremities.
A program which implements this method is available on the appendix of \cite{Belaloui2019}. Unfortunately, using the bidirectional search method we could only obtain results which are less accurate than the one presented above. 
We are investigating different ways to improve the performance of this method notably using sequential search to focus the computational effort on some preferential part of the weave.

\section{Conclusion}
In this paper we have exposed a  TQC scheme in which  the  system is composed of a collection of non-abelian anyons, described by the Fibonacci anyon model.  The Hilbert space of the entire system is the fusion space of the anyons whose basis consists of states characterized by the different ways in which all the anyons can be fused together.

In Sec.\ref{sec:qubit} we have explained how anyons quasi-particles that emerge in two dimensional surfaces -under specific conditions- can be particularly interesting candidate to encode quantum information. Specifically, we have shown how a topological qubit can be designed based on three Fibonacci non-abelian anyons and explained how this implementation provides for leakage free single-qubit braiding gate. Information encoded in such a qubit is topologically protected and is processed by braiding the anyons to methodically manipulate their topological outcome yielding to a fault-tolerant quantum computation.

In Sec.\ref{sec:gate} we have demonstrated how elementary braiding operators can be successively applied on the three anyons composing the topological qubit to construct  braiding gates that approximate any single-qubit quantum gate to a given accuracy.
We presented a leakage free braiding approximation of the Hadamard gate through a thirteen braid long weave with an error of 0.00657. The weave has been constructed using a brute force search method. We have explained the employed algorithms and  shared online the Python codes that implement those algorithms. This codes could be used for reproduction, improvement and further construction of any single-qubit braiding gate with similar or better accuracy if adequate computational power is available.
Furthermore, we are investigating multiple possibilities to improve our accuracy approximating single-qubit gates with braiding either by improving the existing algorithms combined to a larger computational capacities or by exploring more sophisticated braiding gates building methods.
Moreover, it might be interesting to investigate the effects of increasing the number of anyons in each qubit as this will increase the number of elementary exchange operations that are possible within each qubit. Therefore,   with more elementary operations, it may be possible to approximate a desired unitary operation with better accuracy.

\section*{Acknowledgement}
The present study has been achieved based on a team work cooperation with Nacer Eddine Belaloui and Abdellah Tounsi as part of their master dissertations. Numerical programs have been elaborated by N.E. Belaloui. We thank Professor Achour Benslama for fruitful discussions.

\section*{References}
\bibliographystyle{iopart-num}
\bibliography{Bibliography_TQC_v04}

\end{document}